\documentclass[letter]{aa} 
\usepackage[colorlinks=true,     linkcolor=blue, citecolor=blue, filecolor=blue, urlcolor=blue]{hyperref} 

\usepackage{color}
\usepackage{longtable}
\usepackage{multicol}
\usepackage{gensymb}
\usepackage{subfig}
\usepackage{graphicx}
\usepackage{siunitx}

\let\linenumbers\relax
\let\endlinenumbers\relax

\usepackage{txfonts}
\usepackage{color}

\begin{document}

\title{Bar-driven gas redistribution suppresses star formation in spiral galaxies: Evidence from dust lanes in NGC~3351}



\author{K. George\inst{1}\fnmsep\thanks{koshyastro@gmail.com}\and  S. Subramanian\inst{2,3}}


\institute{University Observatory, LMU Faculty of Physics, Scheinerstrasse 1, 81679 Munich, Germany \and Indian Institute of Astrophysics, Koramangala II Block, Bangalore, India \and  Leibniz-Institut fur Astrophysik Potsdam (AIP), An der Sternwarte 16, D-14482 Potsdam, Germany
}

  \abstract{We present observational evidence, based on high-resolution imaging from HST, ALMA, and AstroSat/UVIT, that the redistribution of gas driven by the bar in the face-on spiral galaxy NGC~3351 results in suppressed star formation in its central regions. Dust and molecular gas coexist in galaxies, allowing dust lanes observed in galaxies to be used to probe the distribution of gas. In the central regions of NGC~3351, covered by the stellar bar, dust lanes are visible in the HST $F438W - F814W$ colour map, but surprisingly, these areas lack molecular gas and recent star formation. The inward orientation of the dust lane morphology towards the galaxy’s centre suggests that molecular gas was once present in this region but was redistributed to the centre due to the stellar bar’s action. The direction of the dust lanes therefore indicates the past inflows of gas towards the galaxy centre, with their morphology consistently oriented inwards along the bar. These findings support a scenario in which the stellar bar has efficiently channelled molecular gas into the nucleus, building the central reservoir while suppressing star formation along the bar. }
  

\titlerunning{Bar-induced star formation quenching in NGC~3351}
\authorrunning{K. George\inst{1}}

\maketitle
%

\section{Introduction}

The suppression of star formation in regions occupied by a stellar bar, commonly referred to as bar quenching, can contribute to the shutdown of star formation in star-forming spiral galaxies \citep{Man_2018}. Bar quenching operating in the central regions of barred spiral galaxies in the local Universe has been extensively studied \citep{Masters_2010,Masters_2012,Cheung_2013,Gavazzi_2015,James_2016,Spinoso_2017,Khoperskov_2018,James_2018,Donohoe-Keyes_2019,George_2019,George_2020,Newnham_2020,Percival_2020,George_2021,Geron_2021,Kim_2024,Scaloni_2024,Renu_2025}. However, the physical processes responsible for bar-driven quenching remain poorly understood. This is particularly significant in light of recent \textit{James Webb} Space Telescope (JWST) observations, which have revealed mature bars at redshifts greater than 1, as well as massive galaxies hosting stellar bars at redshifts $\sim$ 4, with kinematic evidence indicating ongoing gas redistribution in one such system at a redshift of 2.467 \citep{Kalita_2026,Guo_2025,Geron_2025,Huang_2025}. Understanding the role of such bars in quenching star formation at high redshifts and the exact process by which this happens is therefore crucial for constraining the global contribution of different quenching mechanisms in star-forming spiral galaxies.

A key open question is how the bar suppresses star formation. One scenario is that the bar redistributes gas within the central region, effectively removing the fuel for star formation from within the bar’s corotation radius (\citealt{Combes_1985}; \citealt{Spinoso_2017}). A second scenario is that enhanced turbulence, driven by shear and shocks generated by the bar, prevents the gas from collapsing into star-forming clouds (\citealt{Tubbs_1982}; \citealt{Reynaud_1998}; \citealt{Verley_2007}; \citealt{Haywood_2016}; \citealt{Khoperskov_2018}). Based on a detailed multiwavelength analysis, \citet{George_2019} demonstrate that bar quenching operates in NGC~3351, where a region encompassing the stellar bar is largely devoid of neutral hydrogen, molecular hydrogen, and ongoing star formation. While the absence of gas in the bar region supports a gas redistribution framework, the exact operation of this mechanism remains elusive. Nonetheless, a multiwavelength study of barred galaxies suggests a gradual evolutionary sequence for bar quenching and outlines a model for this redistribution (\citealt{George_2020}; see also \citealt{Scaloni_2024} and \citealt{Renu_2026}).

In this Letter we explore whether dust lanes can be used as tracers of gas flows in barred galaxies and assess whether their morphology provides insight into the gas redistribution suppressing star formation. Dust grains are formed in the cold interstellar medium from the metals produced by stellar nucleosynthesis and released through stellar winds from evolved stars and supernova explosions; these grains then get mixed with the surrounding gas \citep{Draine_2009}. Since dust is generally co-spatial with cold gas, molecular hydrogen is often found associated with dust lanes. Along star-forming spiral arms, narrow dust lanes are commonly observed because they are compressed in regions where gas is dense, tracing the dense gas structures. However, in the dynamic action of bar-driven torques, gas and dust can react differently, with the gas reacting quickly and the dust lagging \citep{Marshall_2008,Sanchez-Menguiano_2015}. By mapping the distribution and morphology of dust lanes, we can thus trace both the current and past locations of molecular gas within galaxies. We exploited this property to investigate the likely flow of molecular gas in the central region of NGC~3351, particularly within the area influenced by the stellar bar. The dust lanes were identified using an $F438W - F814W$ colour map constructed from high-resolution ($\sim0.1\arcsec$) optical imaging obtained with the \textit{Hubble} Space Telescope (HST). We also utilized contours from Atacama Large Millimeter/submillimeter Array (ALMA) $\text{CO}(J=2{-}1)$ maps to trace the molecular gas distribution, along with AstroSat/UVIT far-ultraviolet (FUV) imaging to probe recent star formation in the galaxy. 

Throughout this Letter we adopt a flat cosmology with $H_{0} = 71,\mathrm{km,s^{-1},Mpc^{-1}}$, $\Omega_{\mathrm{M}} = 0.27$, and $\Omega_{\Lambda} = 0.73$ \citep{Komatsu_2011}.

\section{Data and analysis}

\begin{figure*}
\centering
\includegraphics[width=1.1\textwidth, bb=0 0 500 500]{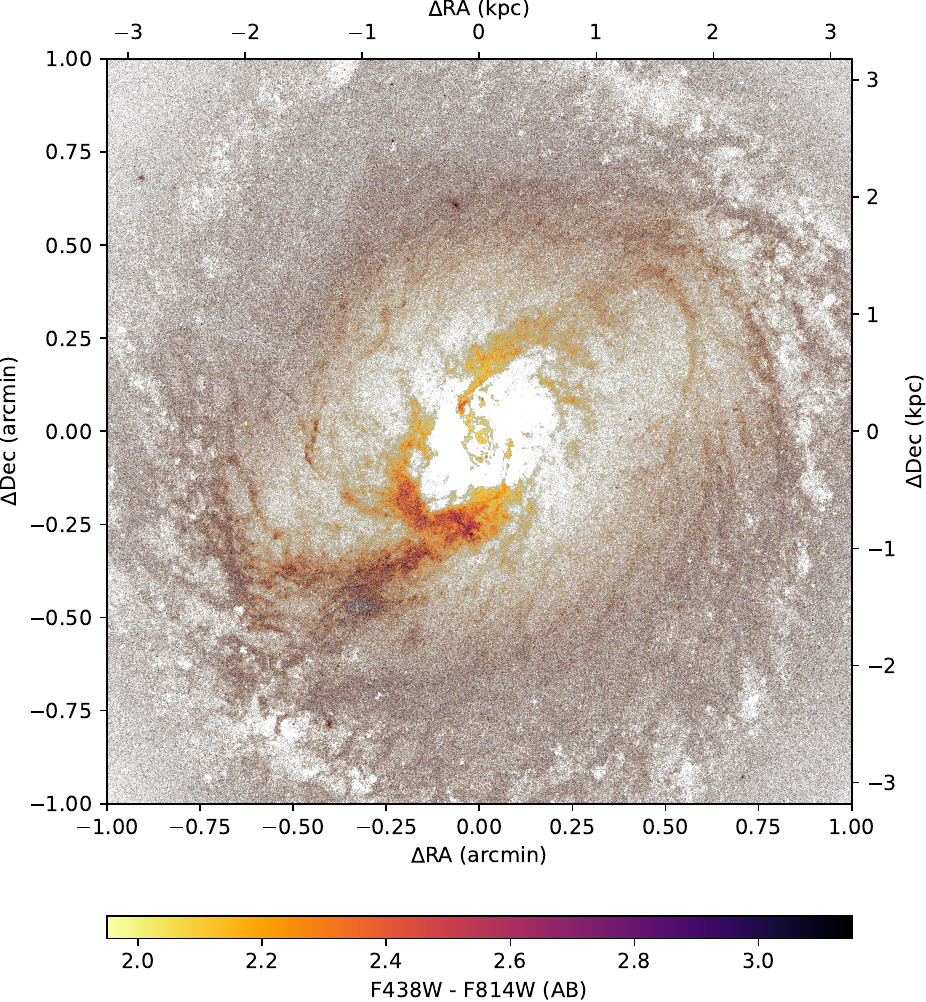}
\caption{High-resolution HST $F438W - F814W$ colour map of the central $\sim2' \times 2'$ region of NGC~3351. Pixels with values in the range 1.95 to 3.15 in $F438W - F814W$ are highlighted; they correspond to colours associated with dust lanes. The displayed field corresponds to a physical size of $\sim6.36\ \mathrm{kpc}$ on each side.
}\label{figure:fig1}
\end{figure*}

NGC~3351\footnote{$\alpha$(J2000) = 10:43:57.7 and $\delta$(J2000) = $+$11:42:14 according to  the NASA/IPAC Extragalactic Database (NED).} (also known as Messier 95) is a nearby (10 $\pm$ 0.4 Mpc; \citealt{Freedman_2001}) early-type barred spiral galaxy (morphology: SBb). The galaxy is nearly face-on (inclination=41$\si{\degree}$, position angle=192$\si{\degree}$) and has a prominent bar. The galaxy has a stellar mass of $\sim$ 10$^{10.4}$ M$_{\odot}$, an HI mass of $\sim$ 10$^{9.2}$ M$_{\odot}$,  an H$_2$ mass of $\sim$ 10$^{9}$ M$_{\odot}$, and an integrated star formation rate of $\sim$ 0.940 M$_{\odot}$/yr \citep{Leroy_2008}. The gas phase metallicity (12 $+$ Log O/H) is 8.60 \citep{Ruyer_2014}. 

We used multiwavelength data of NGC~3351 from the Physics at High Angular resolution in Nearby GalaxieS (PHANGS) survey\footnote{\url{http://www.phangs.org/}} for the analysis presented here. PHANGS is a multiwavelength survey of nearby galaxies carried out with ALMA, the Very Large Array (VLA) HI, HST, AstroSat, JWST, and the Very Large Telescope (VLT) whose goal is to determine the interplay of the small-scale physics of gas and star formation with galactic structure and galaxy evolution. The PHANGS-HST survey \citep{Lee_2022} provides high-resolution, multi-band optical imaging from HST; PHANGS-ALMA \citep{Leroy_2021} provides $\text{CO}(J=2{-}1)$ imaging from ALMA; and PHANGS-AstroSat \citep{Hassani_2024} provides FUV imaging from the UVIT instrument on board AstroSat \citep{Agrawal_2006,Tandon_2017}.

We constructed an HST $F438W - F814W$ colour map of the galaxy, shown in Fig.~\ref{figure:fig1}. The HST $F438W$ and $F814W$ images were photometrically calibrated to the AB system using the zero-point information in the image headers\footnote{https://www.stsci.edu/hst/instrumentation/acs/data-analysis/zeropoints}. These HST filter images were chosen to provide a long wavelength baseline, enabling the most effective sampling of the widest wavelength ranges and maximizing the effect of dust extinction. The colour map was produced from drizzled imaging data available through PHANGS, with fluxes converted to magnitudes using the zero-point calibrations. We display only pixels with colour values in the range 1.95 to 3.15 mag, chosen to isolate strongly reddened regions associated with dust extinction. We verified that varying this threshold by $\pm$0.2 mag does not qualitatively alter the identified dust lane morphology, indicating that our results are robust to the exact choice of colour cut. The resulting colour map can be considered a proxy for the dust distribution in the galaxy.

\begin{figure}
\centering
\includegraphics[width=0.55\textwidth, bb=0 0 500 500]{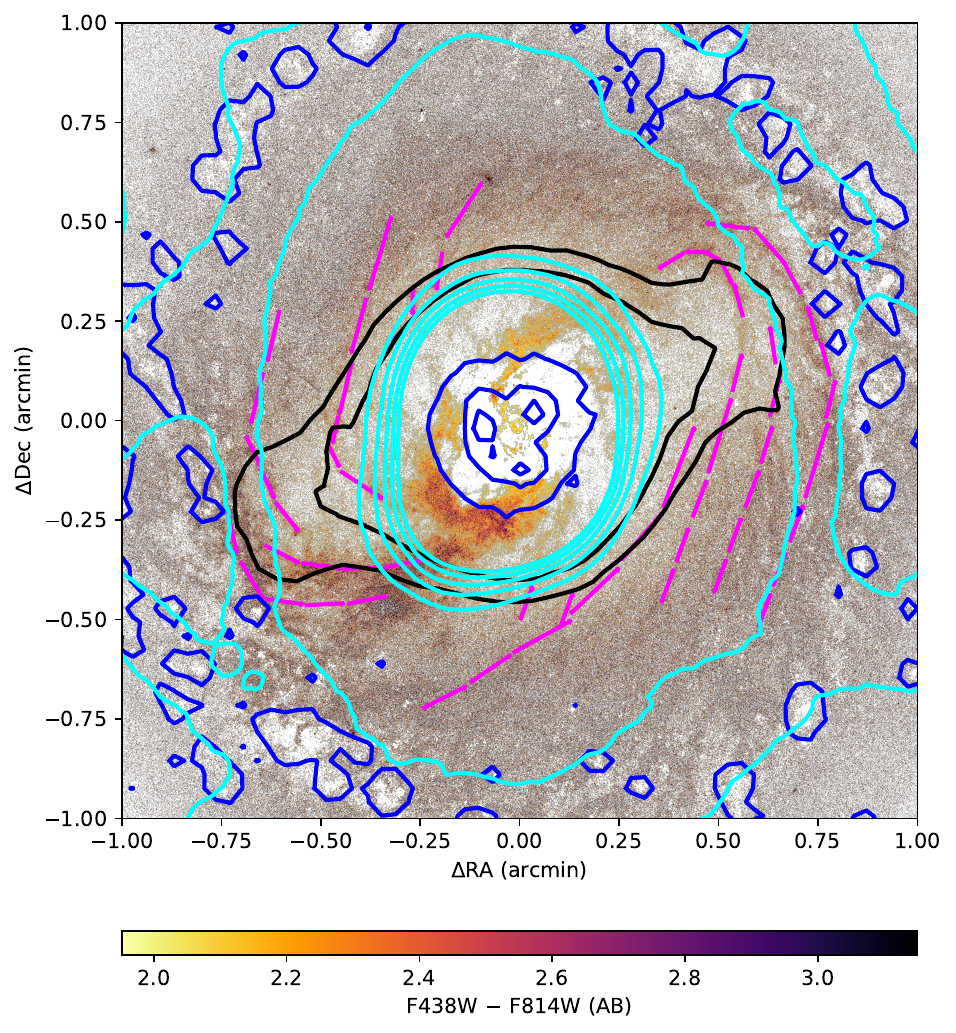}
\caption{ $F438W - F814W$ colour map of NGC~3351 with the stellar bar (black), AstroSat/UVIT FUV (blue), and ALMA $\text{CO}(J=2{-}1)$ map (cyan) contours overlaid. The location of the dust lanes is marked by dashed magenta lines.}\label{figure:fig2}
\end{figure}

Following \citet{George_2019}, we used the \textit{Spitzer} IRAC $3.6~\mu\mathrm{m}$ image as an extinction-free tracer of the evolved stellar population and to delineate the stellar bar \citep{Meidt_2014}. To define the stellar bar, we generated contours from the $3.6~\mu\mathrm{m}$ image and overlaid them (in black), along with contours derived from the ALMA $\text{CO}(J=2{-}1)$ map (in cyan) and the AstroSat/UVIT FUV image (in blue), on the dust map (Fig.~\ref{figure:fig2}). We generated contours from the ALMA CO($J=2-1$) maps convolved to a lower spatial resolution ($\sim 15 \arcsec$) than the native-resolution maps ($\sim 1.5 \arcsec$). Smoothing to a coarser resolution increases the signal-to-noise ratio of large-scale, faint, and extended structures. The two versions exhibit a similar lack of emission along the bar region, while the native-resolution map further resolves structural features in the central regions and outside the bar. We show the five contour levels for ALMA CO($J=2-1$), corresponding to 1.05, 2.10, 3.15, 4.20, and 5.25 $\mathrm{K\ km\,s^{-1}}$. From the infrared image, we estimated the bar length to be $\sim 87\arcsec$ ($\sim 4.2\,\mathrm{kpc}$). These contours are overlaid on a separate figure to maintain the clarity of the dust-lane visualization in Fig.~\ref{figure:fig1}. 
Figure~\ref{figure:fig1} highlights dust lanes in the central region, tracing a spiral-like pattern with varying intensity across the area spanned by the stellar bar. The dust lanes appear to be orientated towards the galaxy centre. This is further illustrated in Fig.~\ref{figure:fig2}, which shows that the bar region — excluding the central nuclear area — lacks detectable molecular gas and ongoing star formation, yet still exhibits prominent dust lanes (marked by magenta lines). Prominent dust lanes are distributed along the stellar bar, and they appear particularly concentrated within it. To quantify the spatial relationship between dust, molecular gas, and star formation, we visually compared their distributions and note a systematic offset, with dust lanes extending across regions where $\text{CO}(J=2{-}1)$ and FUV emission are absent along the region covered by the bar. A full pixel-by-pixel correlation analysis is beyond the scope of this Letter but will be explored in future work.

\section{Discussion}

In the local Universe, star-forming barred spiral galaxies are observed to have a depletion of neutral hydrogen and molecular hydrogen within the central sub-kiloparsec region and at the ends of the bar, specifically within the bar’s corotation radius \citep{Consolandi_2017,George_2020,Newnham_2020}. These regions are characterized by no recent star formation and are predominantly composed of evolved stellar populations \citep{James_2009,James_2015, James_2016,James_2018,Percival_2020,Scaloni_2024,Renu_2025}. Simulations reproduce the observational result, with regions around the bar showing a reduced gas surface density, suppressed star formation, and an older stellar population \citep{Gavazzi_2015,Spinoso_2017,Donohoe-Keyes_2019}. These findings suggest that the central regions of spiral galaxies encompassed by the bar can experience significant star formation suppression.

We examined the phenomenon of star formation suppression in detail, focusing on the stellar bar of NGC~3351, and created a high-resolution optical colour map of the galaxy's central region, which we compared with ALMA $\text{CO}(J=2{-}1)$ and AstroSat/UVIT imaging data. $\text{CO}(J=2{-}1)$ emission is used to trace the bulk molecular gas in galaxies, whereas FUV emission originates from the most recent episodes of star formation. As shown in Fig.~\ref{figure:fig2}, despite the presence of dust lanes in the colour map, ALMA $\text{CO}(J=2{-}1)$ observations fail to detect molecular gas down to the 1$\sigma$ mass sensitivity limits of $\sim$ 2 $\times$ 10$^4$ M$\odot$ at a physical resolution of 100 pc \citep{Leroy_2021}. This region is largely devoid of FUV emission, and therefore recent star formation, indicating bar-driven quenching. In contrast, the central region of the bar contains a substantial amount of molecular hydrogen and exhibits ongoing nuclear star formation. Interestingly, dust lanes are detected not only along the bar but also in regions outside it, yet still within the bar’s corotation radius. In these areas, $\text{CO}(J=2{-}1)$ emissions remain undetected, while dust structures persist. Since dust typically traces cold gas, the presence of dust suggests that gas was once associated with these regions but has since been largely depleted. The dust lanes likely map the pathways of gas redistribution within the bar-influenced central region. Their structural orientations (Fig.~\ref{figure:fig2}) support a scenario in which gas flows along the bar as well as through adjacent regions, driving the efficient redistribution of cold gas across the central environment. This provides observational evidence of gas redistribution driven primarily by bar-induced dynamical processes. The most plausible explanation is that the gas in NGC~3351 is rapidly transported along the bar, which leaves dust behind, creates a gas-deficient region devoid of fuel for further star formation, and thereby suppresses star formation. While a magnetic-field decoupling scenario can explain spatial offsets between dust and gas along the bar as observed in the Milky Way \citep{Marshall_2008} and NGC~1097 \citep{Beck_2005}, the near-complete absence of molecular gas across the wider bar region of NGC 3351 points to a more radical scenario. Here, the prominent dust lanes likely act as structural fossils, tracing the past trajectory of gas that has already been efficiently evacuated and channelled towards the galactic centre.

We investigated recent star formation along the bar of NGC~3351 using H$\alpha$ imaging. H$\alpha$ traces very recent star formation ($\lesssim10$ Myr), while FUV emission traces stars up to $\sim200$ Myr old \citep{Kennicutt_Evans_2012}. Figure~14 of \citet{Emsellem_2022} shows that the H$\alpha$ emission is strongly concentrated in the central regions, with little emission along the stellar bar. A similar distribution is seen in the AstroSat/FUV flux map (Fig.~\ref{figure:fig2}), where the FUV emission is also predominantly centrally concentrated. 

The region where we detect a molecular gas deficiency is also devoid of neutral hydrogen \citep{George_2019}, confirming a near-complete lack of total gas content. We estimated the expected bulk molecular hydrogen mass within a $300 \text{ pc} \times 300 \text{ pc}$ region of the dust lane where ALMA $\text{CO}(J=2{-}1)$ emission was undetected. To isolate the extinction attributable to the dust lane, we subtracted a smooth stellar background from the colour map. The resulting $F438W - F814W$ colour excess was then converted into $A_V$ values using the extinction coefficients from \citet{Schlafly_2011}. We derived a median value of $A_V = 1$ mag within the probed dust lanes to estimate the gas mass where CO emission was undetected. Using the relation between $A_V$ and total hydrogen column density ($N_{\mathrm{H}}$) from \citet{Bohlin_1978}, we derived a bulk molecular hydrogen mass ($M_{\mathrm{H_2}}$) of  $\sim 1.8 \times 10^6 M_\odot$ (assuming R$_V$ = 3.1). This value is well above the $1\sigma$ mass sensitivity limit of $\sim 2 \times 10^4\ M_\odot$ at a physical resolution of 100 pc reported by \citet{Leroy_2021}, confirming a lack of molecular hydrogen even under the assumption of a minimal dust content along the bar region. The apparent anti-correlation between dust and $\text{CO}(J=2{-}1)$ emission in this region might alternatively reflect the occurrence of a diffuse molecular gas phase, which would be more readily detected in the $\text{CO}(J=1{-}0)$ transition due to its lower critical density. However, the $\text{CO}(J=1{-}0)$ map of NGC 3351 presented by \citet{Lee_2026} also reveals a distinct deficit of flux along the bar region, consistent with the morphology observed in $\text{CO}(J=2{-}1)$ data. The evidence discussed above indicates that both the dust-lane morphology and the lack of molecular gas and star formation in the bar region are consequences of recent gas funnelling towards the galactic centre (see also \citealt{Sextl_2026}). Because the dust lanes trace the prior path of this gas, these features confirm that bars redistribute the molecular gas, depleting fuel and ultimately triggering star formation quenching within the bar region.

We note that alternative scenarios may also explain the star formation activity and molecular gas distribution in the nuclear region of NGC~3351. \citet{Leaman_2019} show that while the main bar dust lanes are consistent with bar-driven shocks and gas inflow, the unusual transverse dust lane likely originates from stellar feedback-driven outflows from the circumnuclear star-forming ring. High-resolution imaging observations reveal two central star-forming rings associated with molecular gas traced by $\text{CO}(J=2{-}1)$ and $\text{CO}(J=3{-}2)$ emission \citep{Leroy_2021,Ruiz-Garcia_2024,Sun_2024}. Both rings are likely linked to bar-induced resonances, with the large-scale bar corotation radius located at 2.2 kpc \citep{Ruiz-Garcia_2024,Devereux_1992}. The inner molecular ring may correspond to the inner Lindblad resonance or a decoupled or nested molecular-gas bar \citep{Devereux_1992,Sloshman_2002}. The gas redistribution scenario discussed here may therefore operate alongside these mechanisms, further enhancing the central molecular gas concentration.

Bar-driven gas redistribution can operate efficiently at high redshifts, following the formation of bars in massive galaxies, and could contribute to the shutdown of star formation in their central regions. The evolution of a dust lane morphology along the bar over time, observed through snapshots, provides additional insight into this process and can be studied using dust-sensitive colour maps of barred spiral galaxies at different redshifts. Thanks to wide-field optical and near-infrared imaging of galaxies out to redshifts of $\sim$ 1, which is becoming available from missions such as \textit{Euclid}, such studies are now within reach \citep{Huertas-Company_2025}. Furthermore, rapid gas consumption outside the bar, following the suppression of star formation within the bar region, may contribute to the global quenching of star formation in spiral galaxies.

\section{Summary}

Combining high-resolution HST imaging with ALMA and AstroSat/UVIT observations of NGC 3351, one of the galaxies in the PHANGS 
survey, we identified areas where gas is likely compressed by creating $F438W - F814W$ colour maps and pinpointing dust lanes using a colour threshold. We identify a region, spanning from the outer extent of the bar to the vicinity of the galactic centre, that shows minimal signs of recent star formation in UVIT FUV imaging and completely lacks molecular gas in ALMA $\text{CO}(J=2{-}1)$ maps.
However, prominent dust lanes are present along this region and are orientated towards the galaxy centre, likely tracing past gas inflow driven by the stellar bar. Recently, gas has been efficiently channelled towards the nucleus, fuelling the central molecular gas reservoir and ongoing star formation while creating a cavity depleted of fuel for further star formation. As a result, star formation is strongly suppressed in the bar region. By applying this dust-lane analysis to a larger sample of barred galaxies, we can investigate systems at different stages of the bar-driven gas redistribution and its effect on star formation.

\begin{acknowledgements}
We thank the anonymous referee for the comments, which
improved the scientific content of the paper.
KG acknowledges the support of DLR through the Euclid@LMU program 50QE2304.
SS acknowledges support from the Alexander von Humboldt Foundation.
This work makes use of data from the PHANGS (Physics at High Angular
resolution in Nearby GalaxieS) collaboration, including observations
obtained with the NASA/ESA \textit{Hubble} Space Telescope (PHANGS-HST), the
Atacama Large Millimeter/submillimeter Array (PHANGS-ALMA), and
AstroSat/UVIT (PHANGS-AstroSat). This paper makes use of the following ALMA data:
ADS/JAO.ALMA\#2012.1.00650.S, ADS/JAO.ALMA\#2015.1.00925.S, ADS/JAO.ALMA\#2015.1.00956.S, ADS/JAO.ALMA\#2017.1.00886.L, and ADS/JAO.ALMA\#2018.1.01651.S. ALMA is a partnership of ESO (representing its member states),
NSF (USA), and NINS (Japan), together with NRC (Canada),
MOST and ASIAA (Taiwan), and KASI (Republic of Korea),
in cooperation with the Republic of Chile.
The Joint ALMA Observatory is operated by ESO, AUI/NRAO, and NAOJ.

Based on observations made with the NASA/ESA \textit{Hubble} Space Telescope,
obtained from the Mikulski Archive for Space Telescopes (MAST)
at the Space Telescope Science Institute (STScI), which is operated
by AURA, Inc., under NASA contract NAS~5-26555.

This work also uses observations obtained with AstroSat,
a mission of the Indian Space Research Organisation (ISRO),
archived at the Indian Space Science Data Centre (ISSDC).
\end{acknowledgements}


\begin{appendix}
\counterwithin*{figure}{part}
\stepcounter{part}
\renewcommand{\thefigure}{A.\arabic{figure}}

\section{Bar-driven gas redistribution scenario}

\begin{figure}
\centering
\includegraphics[width=6cm,height=6cm,keepaspectratio]{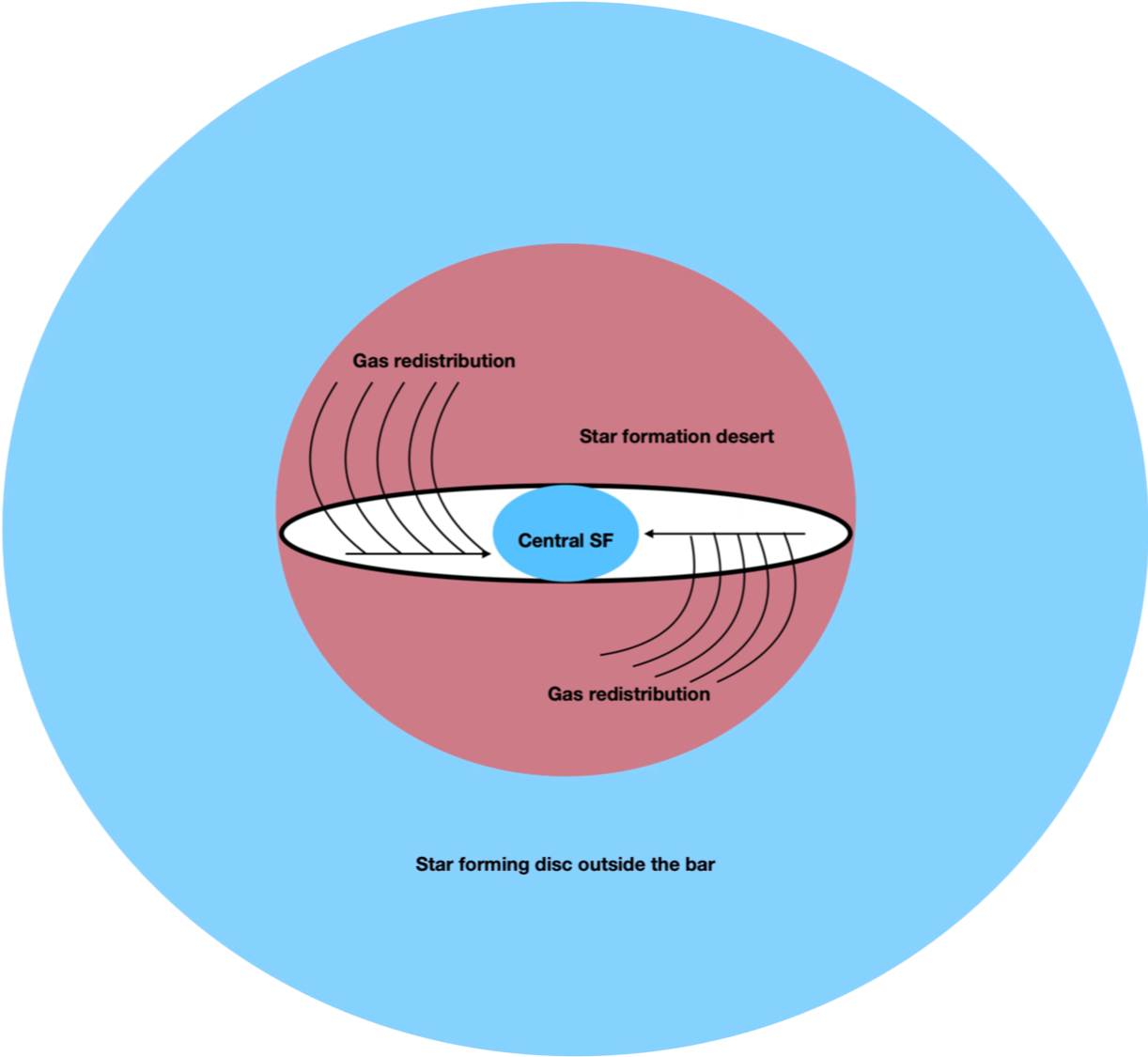}  
\caption{Schematic depiction of gas redistribution in NGC~3351 derived from the dust-lane analysis presented in this work. The bar region and disc are not to scale.}\label{figure:figA2}
\end{figure}

A schematic illustration of the gas redistribution scenario in NGC~3351 based on the dust analysis presented here is shown in Fig.~\ref{figure:figA2}. The central region, corresponding to the extent of the bar (indicated by the black ellipse), is devoid of star formation (shown in red), while star formation is concentrated in the galaxy centre (shown in blue). Gas redistribution occurs throughout the bar region, as indicated by the curved lines, with further transport along the bar towards the centre, where star formation is observed. Additional star formation is present in the disc outside the bar region (also shown in blue). The redistribution of gas creates a central concentration, with a kiloparsec-scale region along the bar that is devoid of gas. In the absence of an external supply of gas, the star formation in the centre will deplete the gas completely, and the galaxy will eventually be devoid of star formation in the bar and the central nuclear region.

\end{appendix}

\end{document}